%% file: main-sigconf.tex
\documentclass[sigconf]{acmart}
\usepackage{graphicx}
\usepackage{algorithm}
\usepackage{rotating}
\usepackage{balance}

\usepackage{tabularx}
\usepackage{subfigure}
\usepackage{bm}
\usepackage{multirow}

\AtBeginDocument{%
  \providecommand\BibTeX{{%
    \normalfont B\kern-0.5em{\scshape i\kern-0.25em b}\kern-0.8em\TeX}}}

\copyrightyear{2021}
\acmYear{2021}
\setcopyright{acmcopyright}
\acmConference[SIGIR '21]{Proceedings of the 44th International ACM SIGIR Conference on Research and Development in Information Retrieval}{July 11--15, 2021}{Virtual Event, Canada}
\acmBooktitle{Proceedings of the 44th International ACM SIGIR Conference on Research and Development in Information Retrieval (SIGIR '21), July 11--15, 2021, Virtual Event, Canada}
\acmPrice{15.00}
\acmDOI{10.1145/3404835.3462843}
\acmISBN{978-1-4503-8037-9/21/07}

\acmSubmissionID{192}

\begin{document}
\fancyhead{}

\title[Learning to Warm Up Cold Item Embeddings for Cold-start Recommendation]{Learning to Warm Up Cold Item Embeddings for Cold-start Recommendation with Meta Scaling and Shifting Networks}


\author{Yongchun Zhu$^{1,2,3}$, Ruobing Xie$^{3}$, Fuzhen Zhuang$^{4,5,*}$, Kaikai Ge$^{3}$, Ying Sun$^{1,2}$, Xu Zhang$^3$, }
\author{Leyu Lin$^3$ and Juan Cao$^{1,2}$}
\affiliation{%
 \institution{$^1$Key Lab of Intelligent Information Processing of Chinese Academy of Sciences (CAS), Institute of Computing Technology, CAS, Beijing 100190, China}} 
\affiliation{%
\institution{$^2$University of Chinese Academy of Sciences, Beijing 100049, China}} 
\affiliation{%
\institution{$^3$WeChat Search Application Department, Tencent, China.}}
\affiliation{%
\institution{$^4$Institute of Artificial Intelligence, Beihang University, Beijing 100191, China.}} 
\affiliation{%
\institution{$^5$SKLSDE, School of Computer Science, Beihang University, Beijing 100191, China.}} 
\affiliation{\{zhuyongchun18s, sunying17g, caojuan\}@ict.ac.cn, \{ruobingxie, kavinge, xuonezhang, goshawklin\}@tencent.com,zhuangfuzhen@buaa.edu.cn}
\thanks{*Fuzhen Zhuang is the corresponding author.}

\renewcommand{\shortauthors}{Y. Zhu et al.}


\begin{abstract}
Recently, embedding techniques have achieved impressive success in recommender systems. However, the embedding techniques are data demanding and suffer from the cold-start problem. Especially, for the cold-start item which only has limited interactions, it is hard to train a reasonable item ID embedding, called cold ID embedding, which is a major challenge for the embedding techniques. 
The cold item ID embedding has two main problems: (1) A gap is existing between the cold ID embedding and the deep model.
(2) Cold ID embedding would be seriously affected by noisy interaction. 
However, most existing methods do not consider both two issues in the cold-start problem, simultaneously.
To address these problems, we adopt two key ideas: (1) Speed up the model fitting for the cold item ID embedding (fast adaptation). (2) Alleviate the influence of noise. Along this line, we propose Meta Scaling and Shifting Networks to generate scaling and shifting functions for each item, respectively. The scaling function can directly transform cold item ID embeddings into warm feature space which can fit the model better, and the shifting function is able to produce stable embeddings from the noisy embeddings. With the two meta networks, we propose Meta Warm Up Framework (MWUF) which learns to warm up cold ID embeddings. Moreover, MWUF is a general framework that can be applied upon various existing deep recommendation models. The proposed model is evaluated on three popular benchmarks, including both recommendation and advertising datasets. The evaluation results demonstrate its superior performance and compatibility.

\end{abstract}

\ccsdesc[500]{Information systems~Recommender systems}

\keywords{Cold-start Recommendation; Item ID Embedding; Warm Up; Meta Network}



\maketitle

\input{Introduction}
\input{Relatedwork}

\input{Model}

\input{Experiment}
\input{Conclusion}

\begin{acks}
The research work is supported by the National Key Research and Development Program of China under Grant No. 2018YFB1004300, the National Natural Science Foundation of China under Grant No. 61773361, U1836206, U1811461. 
\end{acks}

\bibliographystyle{ACM-Reference-Format}
\balance
\bibliography{main-sigconf}

\end{document}

%% file: Introduction.tex
\section{Introduction}
With the explosively growing of personalized online applications, recommender systems have been widely adopted by various online services, including E-commerce, online news, and so on. Traditional collaborative filtering (CF)~\cite{sarwar2001item,koren2009matrix,rendle2009bpr} that can learn user interest and estimate preference from the collected user interactions has shown remarkable performance to build recommender systems. 

Recent studies~\cite{covington2016deep,cheng2016wide,he2017neural,lian2018xdeepfm} have shown that deep networks can further improve the performance of recommender systems. These deep recommendation models adopt the advanced embedding techniques, and can be typically decomposed into two parts: an embedding layer~\cite{pan2019warm} and a deep model. The embedding layer can transform the raw feature into low-dimensional representations (embeddings). Especially, item ID embedding is transformed from an item identifier (item ID), which can be viewed as a latent representation of the specific item. Moreover, it has been widely known in the industry that a well-trained ID embedding can largely improve the recommendation performance~\cite{covington2016deep,cheng2016wide,guo2017deepfm}. Then, all feature embeddings are fed into the deep model to get the final prediction. The deep model can be any structures, e.g., the models learn high-order interactions~\cite{wang2017deep,he2017neural,cheng2019adaptive}, sequential models~\cite{hidasi2015session,zhou2018deep}. These methods have achieved state-of-the-art performance across a wide range of recommendation tasks.



However, these deep recommendation models are data demanding and suffer from the cold-start problem~\cite{volkovs2017dropoutnet,lee2019melu,pan2019warm}. Especially, for the cold-start item which only has limited interactions, it is hard to train a reasonable item ID embedding, called cold ID embedding, which aggravates the cold-start problem of the deep recommendation models. With the cold item ID embedding, it is hard to make satisfying recommendations for the cold item. There are two major problems with the cold item ID embedding.
\begin{itemize}
    \item \textbf{A gap is existing between the cold ID embedding and the deep model.} A small number of items (hot items) account for most samples, while a large number of cold items only have very limited data~\cite{pan2019warm}. The deep model trained with all data would learn much knowledge from the hot items. However, the cold item ID embedding is learned with limited samples of the specific new item, and it is hard for the cold item ID embedding to fit the deep model. Thus, speeding up the model fitting for cold items is important (fast adaptation).
    \item \textbf{The cold item ID embedding would be seriously influenced by noise.} In recommender systems, there are many noisy interactions every day such as wrong clicks. Parameters of item ID embedding are sensitive to the samples of the specific item, while other parameters not. As a result, even small noise can have a serious negative impact on the cold ID embedding generated from limited samples, called noisy embedding~\cite{wang2019sequential}. Hence, it is necessary to alleviate the inﬂuence of noise on cold ID embeddings.
\end{itemize}

Some approaches have been proposed to address the challenges in the cold-start problem. DropoutNet~\cite{volkovs2017dropoutnet} applies dropout to control inputs and utilizes the average embeddings of interacted users to replace the item ID embedding. MetaEmb~\cite{pan2019warm} utilizes a generator to generate a good initial ID embedding from item features for each new item, and the initialization adapts fast when a minimal amount of data is available. Abandoning user and item ID embeddings, MeLU~\cite{lee2019melu} learns a global parameter to initialize the parameter of personalized recommendation models, and the personalized model parameters will be locally updated with limited samples to fast adapt user's preference. Some other methods~\cite{dong2020mamo,lu2020meta} share the similar idea to MeLU. However, all these methods do not consider both two issues in the cold-start problem, simultaneously.

To address the two challenges, we adopt two key ideas, respectively. 1) Recent study~\cite{chen2020esam} indicates that the feature space of cold and warm ID embeddings would be different. To achieve fast adaptation, our first idea is directly transforming the cold ID embeddings into the warm feature space which can fit the model better. 2) Due to the noise in cold ID embeddings, the transformed embeddings are also noisy. To alleviate the influence of noise, the second idea is utilizing the global interacted logs to enhance the ID embeddings.

In this paper, according to the number of interaction samples, we divide the cold-start problem into two phases: cold-start phase (zero sample) and warm-up phase (a few samples). To address the two-phases of the cold-start problem, we propose a general Meta Warm Up Framework (MWUF), which can be applied upon various existing deep recommendation models. The framework consists of a common initial ID embedding and two meta networks. 1) For all new come items, we initialize its embedding with the average embeddings of all existing items, called the commong initial embedding. 2) Meta Scaling Network takes features of an item as input and generates a customized scaling function to transform cold ID embeddings into warmer embeddings, which is able to achieve fast adaptation. 3) With the mean embeddings of global interacted users as input, Meta Shifting Network is able to produce a shifting function that can produce stable embeddings from the noisy embeddings. Following the classical meta-learning paradigm~\cite{munkhdalai2017meta}, we name these two networks “meta network” because they can generate different functions for different items. Note that the proposed MWUF framework is an extra module which can be applied upon various existing models, and only cold items need to pass the MWUF, which means the MWUF has no influence on the hot items. In this paper, we mainly focus on the warm-up phase, while the common initial embedding can also achieve satisfying performance on the cold-start phase.

The main contributions of this work are summarized into three folds:
\begin{itemize}
    \item We propose a general Meta Warm Up Framework (MWUF) to warm up cold ID embeddings. Moreover, MWUF is easy to be applied upon various existing deep recommendation models.
    \item The proposed MWUF is convenient to implement in online recommender systems. Once two meta networks are trained, it is easy to utilize them to transform cold ID embedding to fit the model better.
    \item We perform experiments on three real-world datasets to demonstrate the effectiveness of MWUF. In addition, we apply the MWUF on six deep recommendation models to testify the compatibility of MWUF.
\end{itemize}

%% file: Relatedwork.tex
\section{Related Work}
In this section, we will introduce the related work from three aspects: Cold-start Recommendation, Meta Learning, Sequential Recommendation.

\textbf{Cold-start Recommendation:} Making recommendations for new users or items with limited interaction samples is a challenging problem in recommender systems, also named cold-start problem. According to the number of interaction samples, the cold-start problem can be divided into two phases: cold-start phase (zero sample) and warm-up phase (a few samples).

For the cold-start phase, it is common to use auxiliary information, e.g., user attributes~\cite{seroussi2011personalised,li2019zero}, item attributes~\cite{mo2015image,xi2019modelling,xie2020internal}, knowledge graph~\cite{wang2018ripplenet}, samples in auxiliary domain~\cite{man2017cross}.

The warm-up phase is a dynamic process that gradually improves the recommendation performance with the increase of samples. 
DropoutNet~\cite{volkovs2017dropoutnet}, MetaEmb~\cite{pan2019warm} and MeLU~\cite{lee2019melu} are able to solve the cold-start problem in warm-up phase, and they have been introduced above.
MetaHIN~\cite{lu2020meta} and MAMO~\cite{dong2020mamo} share similar idea with MeLU. 
These methods can be categorized into three groups: (1) MeLU, MAMO, MetaHIN try to personalize the parameters of the deep model. (2) MetaEmb exploits a good pre-trained embedding. (3) DropoutNet learns a more robust item embedding.
Note that the proposed MWUF is completely different from these methods, and does not belong to the three groups. MWUF learns to warm up cold items by using meta networks to predict scaling and shifting functions which can transform the cold ID embeddings to fit the model better.

\textbf{Meta Learning:} Also known as learning to learn, meta-learning intends to learn the general knowledge across similar learning tasks, so as to rapidly adapt to new tasks based on a small number of examples~\cite{vilalta2002perspective}. The work in this stream can be grouped into three clusters, i.e., metric-based~\cite{snell2017prototypical}, optimization-based~\cite{finn2017model}, and parameter-generating approaches~\cite{munkhdalai2017meta}.

Recently, lots of meta learning methods have been proposed for recommendation. The work in~\cite{vartak2017meta} takes a concept of the metric-based meta-learning algorithm in the recommender system to predict whether a user consumes an item or not. $s^2$Meta~\cite{du2019sequential} learns a meta-learner to initialize recommender models of novel scenarios with a few interaction. SML~\cite{zhang2020retrain} utilizes meta learning for the retraining task. Some methods~\cite{pan2019warm,lee2019melu,lu2020meta,dong2020mamo} apply MAML~\cite{finn2017model} into recommender systems. MetaEmb~\cite{pan2019warm} and MeLU~\cite{lee2019melu} are the most related work, which falls into optimization-based methods, while our method falls into the third group which uses one neural network to generate the parameters of another network~\cite{li2019lgm}.

\textbf{Sequential Recommendation:} Sequential recommendation focuses on effectively exploiting the sequential interaction data. There are various sequential recommendation methods: markov chain based~\cite{rendle2010factorizing}, CNN based~\cite{tang2018personalized}, RNN based~\cite{hidasi2015session}, attention based~\cite{zhou2018deep,zhu2020modeling} and mixed models~\cite{ying2018sequential,xi2020neural,xi2021modeling}. Some cold-start methods~\cite{vartak2017meta,lee2019melu,lu2020meta,dong2020mamo} also focus on how to effectively utilize the limited samples. Thus, we consider that the sequential recommendation models can be helpful for the cold-start problem, and we compare cold-start methods with the sequential methods in experiments.

%% file: Model.tex
\begin{figure*}[ht]
\centering
\begin{minipage}[b]{1\linewidth}
\centering
\includegraphics[width=0.85\linewidth]{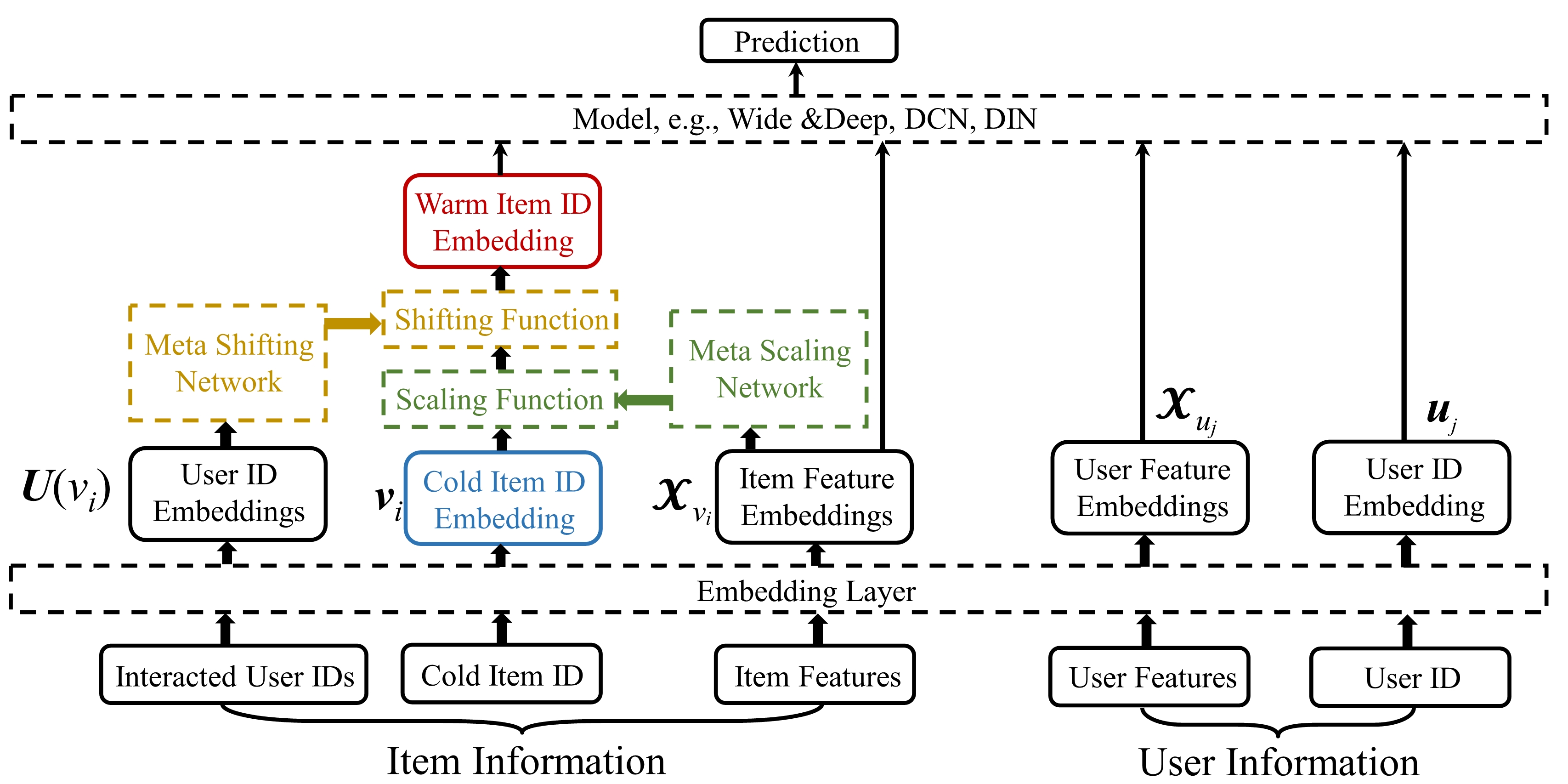}
\end{minipage}
\caption{The proposed Meta Warm Up Framework which consists of a Meta Scaling Network and a Meta Shifting Network to generate scaling and shifting functions, respectively.}\label{fig:network}
\end{figure*} 
\section{Model}

\subsection{Problem Definition}
In this paper, we focus on binary classification recommendation tasks, e.g., CTR, predicting purchasing behavior. Each sample includes a user, an item and a label $y\in\{0, 1\}$. Both user and item contains some features, including user ID, item ID, user age, item category and so on. With the advantage of embedding techniques, these raw features will be transformed into dense vectors, called embeddings. Following~\cite{pan2019warm,zhu2020modeling}, for those categorical inputs (i.e., user IDs and item IDs), we can map each value into a unique embedding. Besides, we can scale embedding vectors by their input values, in order to account for continuous valued inputs. After the embedding layer, we get an item ID embedding $\bm{v}_i$ for the $i$-th item $v_i$, and a set of item feature embeddings $\bm{\mathcal{X}}_{v_i} = \{\bm{x}_{v_i}^1, \cdots, \bm{x}_{v_i}^n\}$, where $\bm{x}_{v_i}^l, l\in[1,n]$ denotes the $l$-th feature of item $v_i$, and $n$ denotes the number of item feature (except item ID). Similarly, for a user $u_j$, we have a user ID embedding $\bm{u}_j$, and a set of user feature embeddings $\bm{\mathcal{X}}_{u_j} = \{\bm{x}_{u_j}^1, \cdots, \bm{x}_{u_j}^m\}$, where $m$ denotes the number of user features. Note that all embeddings $\bm{u},\bm{v},\bm{x}\in \mathbb{R}^{k}$ where $k$ denotes the dimension of embeddings. Therefore, a sample can be denoted as $([\bm{v}_i, \bm{\mathcal{X}}_{v_i}, \bm{u}_j, \bm{\mathcal{X}}_{u_j}], y)$, so we predict $\hat{y}$ by a discriminative function $f(\cdot)$ from the sample:
\begin{equation}
    \hat{y} = f(\bm{v}_i, \bm{\mathcal{X}}_{v_i}, \bm{u}_j, \bm{\mathcal{X}}_{u_j}; \theta),
\end{equation}
where $\theta$ denotes the parameters of the deep model $f(\cdot)$. While the interacted users of an item are useful for prediction~\cite{liu2019real,wu2019dual}, we denote the interacted user set of the item $v_i$ as $U(v_i)$, and $\bm{U}(v_i)$ contains the user ID embeddings of users in $U(v_i)$. It is optimal to feed the models with $U(v_i)$ to further improve the recommendation performance. The Log-loss is often used as the optimization target for binary classification:
\begin{equation}
    \mathcal{L}(\theta, \phi) = -y \log \hat{y} - (1-y) \log (1 - \hat{y}),
\end{equation}
where $\phi$ denotes the parameters of the embedding layer, including $\bm{v}, \bm{u}, \bm{x}$. In the item cold-start problem, a cold item has been interacted by limited users $U(v_i)$. The cold-start phase and warm-up phase can be denoted as $|U(v_i)|=0$ and $|U(v_i)|>0$, respectively. In this paper, we focus on the warm-up phase. 

\subsection{Common Initial ID Embedding}\label{inital}
In recommender systems, a well-trained item ID embedding can largely improve the recommendation performance, and the item ID embedding can be viewed as a latent representation of the specific item. When a new item comes, the item ID embedding would be randomly initialized. However, the randomly initialized ID embedding contains little useful information about the specific item. Thus, with the randomly initialized item embeddings, the recommendation performance is unsatisfying for the cold-start items. 

In addition, some studies~\cite{li2019lgm,shen2018neural} indicate that meta networks that can generate parameters are inherently hard to optimize. Meanwhile, we find the randomly initialized ID embeddings could make it difficult to train meta networks well. Therefore, to train the two meta networks well, it is essential to look for a good initialization approach. Some studies~\cite{finn2017model} of meta Learning think the common knowledge is useful, and it is easy to fit specific tasks by fine-tuning. Inspired by this idea, we want to utilize a common initial embedding, and it is intuitive to take the average embeddings of all existing items to initialize ID embeddings of new items. For each new item, we initialize its embedding with the common initial ID embedding. It is obvious that the common initialization method is better than random initialization. In the warm-up phase, with some interactions, it is convenient to train the ID embedding to better represent the specific item. Note that the common initial embedding is not only good for the training of two meta networks, but also helpful for the cold-start phase (no sample).


\subsection{Two Meta Networks}
A good initial ID embedding is helpful for speeding up the fitting process. However, with a small number of steps to update, the cold ID embeddings are still hard to fit the model. To warm up cold item ID embeddings, we propose two meta networks to generate scaling and shifting functions to directly transform the cold embeddings into warmer and more stable embeddings that fit the model better. The model is shown in Figure~\ref{fig:network}.

\textbf{Meta Scaling Network:} 
A recent study~\cite{chen2020esam} indicates that the feature space of cold and warm ID embeddings would be different. For each item, there is a relationship between the cold ID embedding and the warm ID embedding (warm ID embedding denotes the embedding of the items with relatively sufficient interactions). Thus, we want to bridge the cold and warm ID embeddings. An interesting finding is that for similar items, they could have similar warm ID embeddings. Thus, for similar items, it is intuitive that the relationships between cold ID embeddings and warm ID embeddings would be similar, too.
To measure similarity, the most reliable way is learning from interactions. However, the cold items have too limited interactions. Therefore, the similarity learned from interactions is unreliable for the cold items.

Fortunately, the features of items are available and relatively stable. Besides, the item features are able to measure the similarity between items~\cite{pan2019warm,li2019zero}. Thus, the relationship between the cold and warm ID embedding can be related to the item features. Inspired by this finding, we propose Meta Scaling Network takes item feature embeddings $\bm{\mathcal{X}}_{v_i}$ as input to generate a scaling function, which can warm up the cold ID embeddings. We formulate Meta Scaling Network as:
\begin{equation}
    \bm{\tau}^{scale}_{v_i} = h(\bm{\mathcal{X}}_{v_i};w_{scale}), \ \bm{\tau}^{scale} \in \mathbb{R}^{k},
\end{equation}
where $w_{scale}$ is the parameters of $h(\cdot)$. Thus, the warm ID embedding transformed from cold ID embedding of item $v_i$ can be denoted as $\bm{v}_i^{warm} = \bm{v}_i \odot \bm{\tau}^{scale}_{v_i}$. The scaling function can be seen as a kind of feature transformation, which transforms the cold ID embedding into the warm feature space. 

\textbf{Meta Shifting Network:} The noise has a negative impact on the cold ID embedding learned from limited samples. Thus, the warm ID embedding transformed from the noisy cold embedding $\bm{v}_i$ is also noisy. Recent study~\cite{liu2019real} indicates that utilizing the mean of interacted users' embeddings to present the item could alleviate the influence of outlier (wrong click). Thus, a direct way is to exploit the global interacted users $U(v_i)$ of the item to enhance the ID embedding. Hence, with the user embedding set $\bm{U}(v_i)$ as input, the Meta Shifting Network can produce a shifting function to make the ID embeddings more stable, and it can be formulated as:
\begin{equation}
    \bm{\tau}^{shift}_{v_i} = g(\mathcal{G}(\bm{U}(v_i));w_{shift}), \ \bm{\tau}^{shift} \in \mathbb{R}^{k},
\end{equation}
where $w_{shift}$ is the parameters of $g(\cdot)$. The size of the interacted user set $U(v_i)$ would be different, so we utilize a function $\mathcal{G}(\cdot)$ to aggregate $\bm{U}(v_i)$. In this paper, we take a simple method unweighted mean as $\mathcal{G}(\cdot)$. Finally, the enhanced warm ID embedding can be denoted as:
\begin{equation}
    \bm{v}_i^{warm} = \bm{v}_i \odot \bm{\tau}^{scale}_{v_i} + \bm{\tau}^{shift}_{v_i}.\label{eq:warm}
\end{equation}
The shifting function can be seen as a way of combining the item ID embedding and its neighbors (users) into a more stable item representation. In the Equation~(\ref{eq:warm}), the cold ID embedding is scaled by $\bm{\tau}^{scale}_{v_i}$ and then shifted by $\bm{\tau}^{shift}_{v_i}$, which is also the origin of the name.
Note that both $\bm{\tau}^{scale}_{v_i}$ and $\bm{\tau}^{shift}_{v_i}$ vary from item to item. Thus, with our method, the warm-up procedure can capture the characteristics of items. The parameters of scaling and shifting functions are generated by the two high-level networks, so we called them meta networks.

\subsection{Overall Procedure}
The overall training procedure is divided into two steps. The first step is to train a recommendation model with all data, which contains parameters  $\theta$ and $\phi$. Though this model has good performance for the overall recommendation, it is unsatisfying to recommend new items. The main reason is that the cold ID embedding cannot fit the model well. In real-world recommender systems, we can directly utilize their unique pre-trained model as $f_\theta$. This step is not our concern.


\begin{algorithm} [t] 
    \caption{Meta Warm Up Framework (MWUF).}\label{alg}
    
    \flushleft{\textbf{Input:} $f_\theta$: A pre-trained model.
    
    \textbf{Input:} $h_{w_{scale}}, \ g_{w_{shift}}$: Two meta networks.
    
    \textbf{Input:} $\phi^{new}_{id}$: An initialized item ID embedding layer.
    
    \textbf{Input:} $\mathcal{D}$: A dataset sorted by timestamp.
    \begin{enumerate}
        \item randomly initialize $h_{w_{scale}}, \ g_{w_{shift}}$.
        \item \textbf{while} not converge \textbf{do}:
        \item \quad Sample batch of samples $\mathcal{B}$ from $\mathcal{D}$
        \item \quad Obtain cold ID embeddings $\bm{v}$ of items in $\mathcal{B}$
        \item \quad Obtain warm ID embeddings $\bm{v}^{warm}$ by Equation(\ref{eq:warm})
        \item \quad Evaluate $\mathcal{L}^{cold}$ with $\bm{v}$
        \item \quad Update $\phi^{new}_{id}$ by minimizing $\mathcal{L}^{cold}$
        \item \quad Evaluate $\mathcal{L}^{warm}$ with $\bm{v}^{warm}$
        \item \quad Update $h_{w_{scale}}, \ g_{w_{shift}}$ by minimizing $\mathcal{L}^{warm}$
        \item \textbf{end while}
    \end{enumerate}
    }
\end{algorithm}

The second step is to train the two meta networks to warm up the cold ID embeddings. To avoid disturbing the recommendation of old items, we leave the parameters of $\theta$ and $\phi$ fixed. To train these two meta networks, we need to utilize limited samples of the old items to simulate the cold-start process. The old items denote the existing items which have relatively sufficient interactions.

We denote the pre-trained item ID embedding layer as $\phi^{old}_{id} \in \phi$, including $\bm{v}$. To simulate the cold-start process, we create a new ID embedding layer parameterized by $\phi^{new}_{id}$, including all item embeddings $\hat{\bm{v}}$. 
We utilize the mean vector of the pre-trained item ID embedding $\bm{v}$ to initialize all item ID embeddings $\hat{\bm{v}}$ in $\phi^{new}_{id}$. In other words, all old items are viewed as new items in this training procedure. Then, we calculate the prediction $\hat{y}^{cold} = f(\hat{\bm{v}}_i, \bm{\mathcal{X}}_{v_i}, \bm{u}_j, \bm{\mathcal{X}}_{u_j}; \theta)$ with the cold ID embeddings $\hat{\bm{v}}_i$, and obtain the cold loss $\mathcal{L}^{cold}$. Meanwhile, we obtain the warm ID embedding $\hat{\bm{v}}^{warm}_i$ with the two meta networks by Equation~(\ref{eq:warm}). With $\hat{\bm{v}}^{warm}_i$, we can get another prediction $\hat{y}^{warm} = f(\hat{\bm{v}}^{warm}_i, \bm{\mathcal{X}}_{v_i}, \bm{u}_j, \bm{\mathcal{X}}_{u_j}; \theta)$, and the warm loss $\mathcal{L}^{warm}$. Then, we optimize the new ID embedding layer by minimizing $\mathcal{L}^{cold}$, and optimize the two meta networks by minimizing $\mathcal{L}^{warm}$:


\begin{equation}
    \begin{split}
        \min_{\phi^{new}_{id}} \mathcal{L}^{cold}, \quad \min_{w^{scale}, w^{shift}} \mathcal{L}^{warm}.\label{eq:loss}
    \end{split}
\end{equation}

Finally, we come to our training algorithm, which can update the parameters by stochastic gradient descent in a mini-batch manner, see Algorithm~\ref{alg}. In the training procedure, we can find that the base model and the two meta networks are updated, respectively. In other words, the proposed MWUF is model-agnostic. Thus, MWUF can be applied upon various base models. 

Note that the proposed MWUF can not only be trained with offline data set, but also be trained online with minor modifications by using the emerging new IDs as the training examples. When a new item comes, the cold ID embedding is initialized with the common initial ID embedding. When the new item is interacted with some users, the interactions would be utilized to train the base model ($\theta, \phi$) and the meta networks with Equation~(\ref{eq:loss}), respectively. Note that the base model is learned with $\mathcal{L}^{cold}$, while the meta networks is learned with $\mathcal{L}^{warm}$. Obviously, the training of the meta networks will not influence the base model. Finally, in the test stage, we directly utilize the warm ID embedding $\bm{v}^{warm}$ for prediction.

%% file: Experiment.tex

\section{Experiments}
In this section, we conduct experiments with the aim of answering the following research questions:
\begin{itemize}
    \item[\textbf{RQ1}] How do different methods (popular CF models, sequential methods, cold-start methods) perform in the cold-start setting?
    \item[\textbf{RQ2}] How does MWUF upon various deep recommendation models perform?
    \item[\textbf{RQ3}] What are the effects of Initialization, Meta Scaling, and Meta Shifting Networks in our proposed MWUF?
    \item[\textbf{RQ4}] Can the initialization methods (MetaEmb, MWUF) alleviate the cold-start problem?
\end{itemize}

\subsection{Dataset}
We evaluate the proposed approach on three datasets:

\textbf{MovieLens-1M\footnote{http://www.grouplens.org/datasets/movielens/}:} It is one of the most well-known benchmark dataset. The data consists of 1 million movie ranking instances over thousands of movies and users. Each movie has features including its title, year of release, and genres. Titles and genres are lists of tokens. Each user has features including the user’s ID, age, gender, and occupation. We transform ratings into binary (The ratings at least 4 are turned into 1 and the others are turned into 0).

\textbf{Taobao Display Ad Click\footnote{https://tianchi.aliyun.com/dataset/dataDetail?dataId=56}:} It contains 1,140,000 users from the website of Taobao\footnote{https://www.taobao.com/} for 8 days of ad display / click logs (26 million records). Each ad can be seen as an item in our paper, with features including its ad ID, category ID, campaign ID, brand ID, Advertiser ID. Each user has 9 categorical attributes: user ID, Micro group ID, cms\_group\_id, gender, age, consumption grade, shopping depth, occupation, city level.

\textbf{CIKM2019 EComm AI\footnote{https://tianchi.aliyun.com/competition/entrance/231721/introduction}:} It is an E-commerce recommendation dataset released by Alibaba Group. Each instance is made up by an item, a user and a behavior label ('pv', 'buy', 'cart', 'fav'). Each item has 4 categorical attributes: item ID, item category, shop ID, brand ID, and each user has 4 categorical attributes: user ID, gender, age, purchasing power. Finally, we transform the behavior label into binary (1/0 indicate whether a user has bought an item or not) to fit the problem setting.

\begin{table*}[pt]
  \centering
  \caption{Model comparison on three datasets. All the lines calculate RelaImpr by comparing with Wide \& Deep on each dataset respectively. We record the mean results over ten runs. * indicate $p \le 0.05$, paired t-test of MWUF vs. the best baselines on AUC.}
    \begin{tabular}{cc||cc||cc||cc||cc}
    \toprule
    \multirow{15}[10]{*}{\begin{sideways}MovieLens-1M\end{sideways}} & \multirow{2}[2]{*}{Methods} & \multicolumn{2}{c||}{cold} & \multicolumn{2}{c||}{warm-a} & \multicolumn{2}{c||}{warm-b} & \multicolumn{2}{c}{warm-c} \\
          &       & AUC   & RelaImpr & AUC   & RelaImpr & AUC   & RelaImpr & AUC   & RelaImpr \\
\cmidrule{2-10}          & FM    & 0.5250  & -74.2\% & 0.5296  & -70.0\% & 0.5406  & -67.4\% & 0.5525  & -65.5\% \\
          & Wide \& Deep    & 0.5968  & 0.0\% & 0.5987  & 0.0\% & 0.6247  & 0.0\% & 0.6523  & 0.0\% \\
          & PNN   & 0.5920  & -5.0\% & 0.5949  & -3.9\% & 0.6240  & -0.6\% & 0.6493  & -2.0\% \\
          & DCN   & 0.6027  & 6.1\% & 0.6046  & 6.0\% & 0.6312  & 5.2\% & 0.6612  & 5.8\% \\
          & AFN   & 0.5862  & -11.0\% & 0.5966  & -2.1\% & 0.6541  & 23.6\% & 0.7029  & 33.2\% \\
\cmidrule{2-10}          & DropoutNet & 0.6432  & 47.9\% & 0.6499  & 51.9\% & 0.6565  & 25.5\% & 0.6628  & 6.9\% \\
          & MeLU  & \textbf{0.6434 } & \textbf{48.1\%} & 0.6400  & 41.8\% & 0.6643  & 31.8\% & 0.6760  & 15.6\% \\
          & MetaEmb & 0.6369  & 41.4\% & 0.6349  & 36.7\% & 0.6686  & 35.2\% & 0.6983  & 30.2\% \\
\cmidrule{2-10}          & GRU4Rec & 0.6032  & 6.6\% & 0.6784  & 80.7\% & 0.6799  & 44.3\% & 0.6872  & 22.9\% \\
          & DIN   & 0.6077  & 11.3\% & 0.6632  & 65.3\% & 0.6747  & 40.1\% & 0.6895  & 24.4\% \\
\cmidrule{2-10}          & MWUF(Wide \& Deep)  & 0.6339  & 38.3\% & 0.6569  & 59.0\% & 0.6999  & 60.3\% & 0.7273  & 49.2\% \\
          & MWUF(GRU4Rec) & 0.6298  & 34.1\% & \textbf{0.6962*} & \textbf{98.8\%} & 0.7033  & 63.0\% & 0.7160  & 41.8\% \\
          & MWUF(AFN) & 0.6370  & 41.5\% & 0.6913  & 93.8\% & \textbf{0.7261*} & \textbf{81.3\%} & \textbf{0.7447*} & \textbf{60.7\%} \\
    \midrule
    \multirow{15}[10]{*}{\begin{sideways}Taobao Display AD\end{sideways}} & \multirow{2}[2]{*}{Methods} & \multicolumn{2}{c||}{cold} & \multicolumn{2}{c||}{warm-a} & \multicolumn{2}{c||}{warm-b} & \multicolumn{2}{c}{warm-c} \\
          &       & AUC   & RelaImpr & AUC   & RelaImpr & AUC   & RelaImpr & AUC   & RelaImpr \\
\cmidrule{2-10}          & FM    & 0.5020  & -91.4\% & 0.5064  & -80.4\% & 0.5096  & -79.7\% & 0.5126  & -80.1\% \\
          & Wide \& Deep    & 0.5233  & 0.0\% & 0.5326  & 0.0\% & 0.5474  & 0.0\% & 0.5634  & 0.0\% \\
          & PNN   & 0.5156  & -33.0\% & 0.5264  & -19.0\% & 0.5401  & -15.4\% & 0.5515  & -18.8\% \\
          & DCN   & 0.5206  & -11.6\% & 0.5296  & -9.2\% & 0.5446  & -5.9\% & 0.5574  & -9.5\% \\
          & AFN   & 0.5241  & 3.4\% & 0.5465  & 42.6\% & 0.5703  & 48.3\% & 0.5810  & 27.8\% \\
\cmidrule{2-10}          & DropoutNet & 0.5214  & -8.2\% & 0.5318  & -2.5\% & 0.5515  & 8.6\% & 0.5655  & 3.3\% \\
          & MeLU  & 0.5226  & -3.0\% & 0.5223  & -31.6\% & 0.5256  & -46.0\% & 0.5360  & -43.2\% \\
          & MetaEmb & 0.5250  & 7.3\% & 0.5345  & 5.8\% & 0.5506  & 6.8\% & 0.5633  & -0.2\% \\
\cmidrule{2-10}          & GRU4Rec & 0.5211  & -9.4\% & 0.5271  & -16.9\% & 0.5394  & -16.9\% & 0.5494  & -22.1\% \\
          & DIN   & 0.5197  & -15.5\% & 0.5281  & -13.8\% & 0.5448  & -5.5\% & 0.5580  & -8.5\% \\
\cmidrule{2-10}          & MWUF(Wide \& Deep)   & 0.5255  & 9.4\% & 0.5718  & 120.2\% & 0.5890  & 87.8\% & 0.5919  & 45.0\% \\
          & MWUF(GRU4Rec) & 0.5208  & -10.7\% & 0.5626  & 92.0\% & 0.5842  & 77.6\% & 0.5867  & 36.8\% \\
          & MWUF(AFN) & \textbf{0.5302 } & \textbf{29.6\%} & \textbf{0.5872*} & \textbf{167.5\%} & \textbf{0.5958*} & \textbf{102.1\%} & \textbf{0.5965*} & \textbf{45.3\%} \\
    \midrule
    \multirow{15}[10]{*}{\begin{sideways}CIKM2019\end{sideways}} & \multirow{2}[2]{*}{Methods} & \multicolumn{2}{c||}{cold} & \multicolumn{2}{c||}{warm-a} & \multicolumn{2}{c||}{warm-b} & \multicolumn{2}{c}{warm-c} \\
          &       & AUC   & RelaImpr & AUC   & RelaImpr & AUC   & RelaImpr & AUC   & RelaImpr \\
\cmidrule{2-10}          & FM    & 0.5386  & -73.9\% & 0.5420  & -74.7\% & 0.5510  & -70.5\% & 0.5598  & -70.2\% \\
          & Wide \& Deep    & 0.6479  & 0.0\% & 0.6657  & 0.0\% & 0.6730  & 0.0\% & 0.7010  & 0.0\% \\
          & PNN   & 0.6512  & 2.2\% & 0.6665  & 0.5\% & 0.6723  & -0.4\% & 0.7022  & 0.6\% \\
          & DCN   & 0.6596  & 7.9\% & 0.6710  & 3.2\% & 0.6771  & 2.4\% & 0.7008  & -0.1\% \\
          & AFN   & 0.6693  & 14.5\% & 0.6909  & 15.2\% & 0.7038  & 17.8\% & 0.7363  & 17.6\% \\
\cmidrule{2-10}          & DropoutNet & 0.6555  & 5.1\% & 0.6719  & 3.7\% & 0.6854  & 7.2\% & 0.6980  & -1.5\% \\
          & MeLU  & 0.6595  & 7.8\% & 0.6697  & 2.4\% & 0.6762  & 1.8\% & 0.7032  & 1.1\% \\
          & MetaEmb & 0.6642  & 11.0\% & 0.6746  & 5.4\% & 0.6795  & 3.8\% & 0.7095  & 4.2\% \\
\cmidrule{2-10}          & GRU4Rec & 0.6378  & -6.8\% & 0.7266  & 36.8\% & 0.7337  & 35.1\% & 0.7635  & 31.1\% \\
          & DIN   & 0.6428  & -3.4\% & 0.7168  & 30.8\% & 0.7250  & 30.1\% & 0.7545  & 26.6\% \\
\cmidrule{2-10}          & MWUF(Wide \& Deep)   & 0.6637  & 10.7\% & 0.6855  & 11.9\% & 0.7097  & 21.2\% & 0.7236  & 11.2\% \\
          & MWUF(GRU4Rec) & 0.6540  & 4.1\% & \textbf{0.7381*} & \textbf{43.7\%} & \textbf{0.7404 } & \textbf{39.0\%} & \textbf{0.7672 } & \textbf{32.9\%} \\
          & MWUF(AFN) & \textbf{0.6741 } & \textbf{17.7\%} & 0.7126  & 28.3\% & 0.7320  & 34.1\% & 0.7556  & 27.2\% \\
    \bottomrule
    \end{tabular}%
  \label{tab:results}%
\end{table*}%

\subsection{Baselines}
We categorize our baselines into three groups according to their approaches. The first group includes the popular CF methods designed for the overall recommendation. (1) FM~\cite{rendle2010factorization} can capture high-order interaction information cross features. For efficiency, we use the same embedding vectors for the first- and the second-order components. (2) Wide \& Deep~\cite{cheng2016wide} is a deep recommendation model combining a (wide) linear channel with a (deep) nonlinear channel, which has been widely adopted in industry. (3) PNN~\cite{qu2016product} uses a product layer to capture interactive patterns between inter-field categories.
(4) DCN~\cite{wang2017deep} efficiently captures feature interactions of bounded degrees in an explicit fashion.
(5) AFN~\cite{cheng2019adaptive} is a most recent FM-based~\cite{rendle2010factorization} method which can learn arbitrary-order cross features adaptively from data.

The second group includes state-of-the-art methods for the cold-start problem.
    (1) DropoutNet~\cite{volkovs2017dropoutnet} is a popular cold-start method which applies dropout to control input, and exploits the average representations of interacted items/users to enhance the representations of users/items.
    (2) MeLU~\cite{lee2019melu} learns a general model from a meta-learning perspective, and the general model can adapt fast to new users/items.
    (3) MetaEmb~\cite{pan2019warm} is the most related method which also focuses on the warm-up phase. It trains a embedding generator from a meta-learning perspective, and utilizes the generated meta-embeddings to initialize ID embeddings of cold items.

The methods in the third group exploit historical interaction information. Most sequential methods focus on the user's sequential interactions. In this paper, we apply the sequential methods into the item side.
(1) GRU4Rec~\cite{hidasi2015session} uses RNNs to model user sequences for the session-based recommendation.
(2) DIN~\cite{zhou2018deep} exploits the mechanism of attention to activate related user behaviors.

The proposed MWUF is a general framework that can be applied upon various models, which is denoted as MWUF(model name), e.g., MWUF(Wide \& Deep).

\begin{figure*}[t]
\centering
\begin{minipage}[b]{1\linewidth}
\centering
\includegraphics[width=0.95\linewidth]{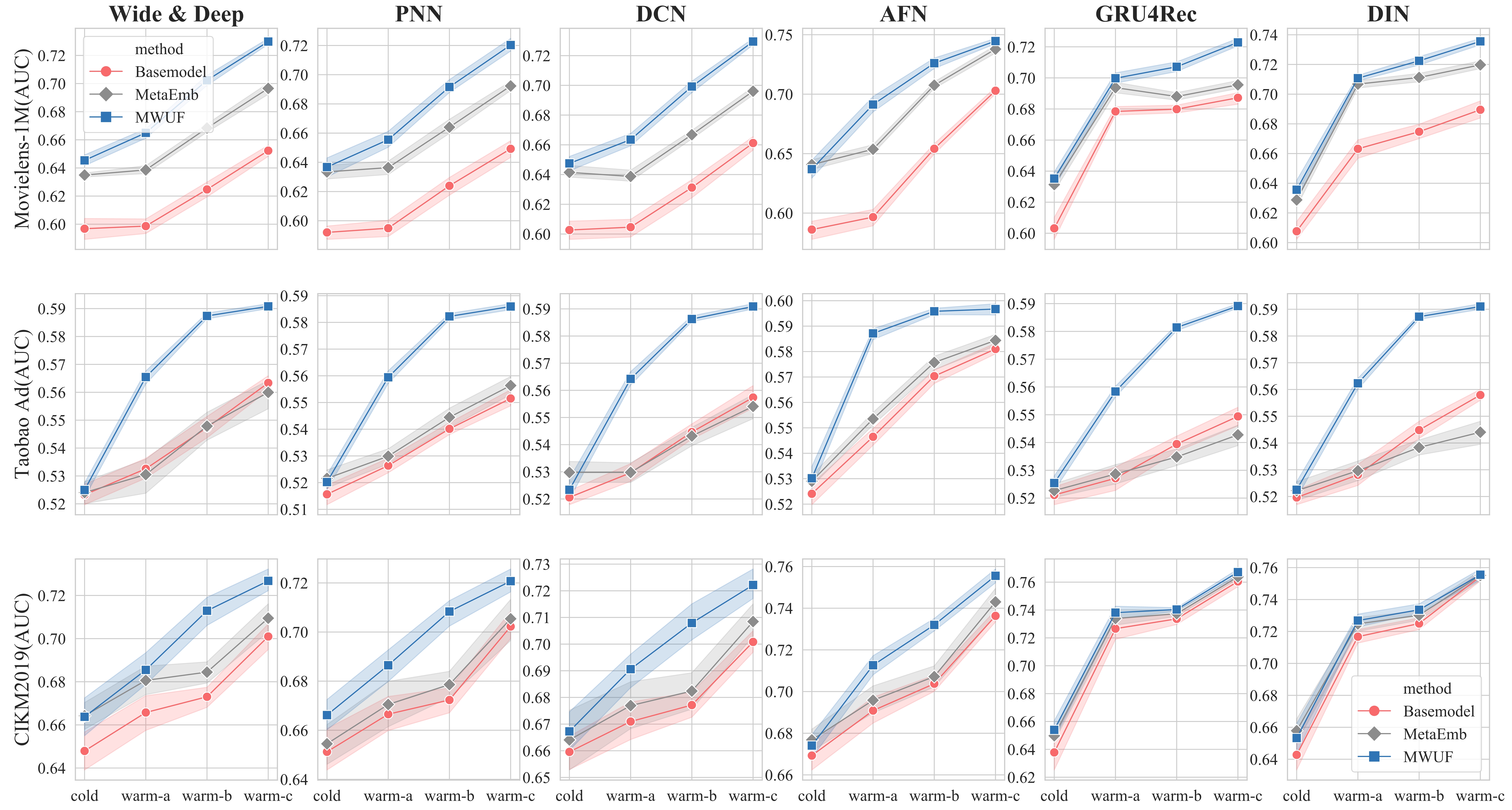}
\end{minipage}
\caption{Performance on three datasets, over six popular base models. The solid lines are averaged scores and the bands are standard deviations over ten runs.}\label{fig:generalization}
\end{figure*}

\subsection{Experimental Settings}
\textbf{Dataset splits.} To evaluate the recommendation performance in both cold-start and warm-up phases, we conduct the experiments by splitting the datasets following~\cite{pan2019warm}. First, we group the items by their sizes:
\begin{itemize}
    \item Old items: The items whose number of labeled instances is larger than a threshold $N$. We use $N$ of 200, 2000, and 350 for MovieLens-1M, Taobao AD data, CIKM2019 data.
    \item New items: Due to the warm-up phase is a dynamic process, we divide the new items into 3 warm-up phases, and each phase contains K samples for each item. Thus, we use the items whose number is less than $N$ and larger than $3 \times K$ as the new items. The samples of each new item are sorted by timestamp, and we use the first $3 \times K$ as 3 warm-up phases denoted as warm-a, -b, -c while the rest as a test set. For the three datasets, K is set as 20, 500, 50.
\end{itemize}
Note that the ratio of new items to old items is approximately 8:2, which is similar to the definition of long-tail items~\cite{chen2020esam}.

\textbf{Implementation  Details.} For a fair comparison, we use the same setting for all methods. The MLP in these models uses the same structure with three dense layers (hidden units 64). The two Meta Networks have the same structure with two dense layers (hidden units 16). The dimensionality of embedding vectors of each input field is fixed to 16 for all experiments. In addition, we set learning rate of 0.001. Training is done through stochastic gradient descent over shuffled mini-batches with the Adam~\cite{kingma2014adam} update rule. For all methods, we set mini-batch size of 256.
Following~\cite{pan2019warm}, for all methods, the experiments are done with the following steps:
\begin{itemize}
    \item[1.] Pre-train the base model with the data of old items.
    \item[2.] Train extra modules or initialize the new item ID.
    \item[3.] Compute evaluation metrics on the warm-a set;
    \item[4.] Update the embeddings of new items IDs with warm-a set and compute evaluation metrics on the warm-b set;
    \item[5.] Update the embeddings of new items IDs with warm-b set and compute evaluation metrics on the warm-c set;
    \item[6.] Update the embeddings of new items IDs with warm-c set and compute evaluation metrics on the test set;
\end{itemize}
Note that only for MetaEmb and our methods, the second step is needed. For MeLU and MWUF, the meta learner will be updated in steps 4, 5, 6. We denote the steps 3-6 as cold, warm-a, warm-b, warm-c phases. Following\cite{pan2019warm}, for all training steps, we update parameters for 1 epoch.

\textbf{Evaluation metrics.} For binary classification tasks, AUC is a widely used metric~\cite{fawcett2006introduction}. It measures the goodness of order by ranking all the items with prediction, including intra-user and inter-user orders. Besides, AUC is a common metric for both recommendation~\cite{vartak2017meta} and advertising~\cite{pan2019warm,zhou2018deep}. Thus, following the cold-start work~\cite{vartak2017meta,pan2019warm}, we adopt AUC as the main metric.
In addition, we follow~\cite{williams1989learning,zhou2018deep} to introduce RelaImpr metric to measure relative improvement over models. For a random guesser, the value of AUC is 0.5. Hence, RelaImpr is defined as below:
\begin{equation}
    \text{RelaImp} = (\frac{\text{AUC(measured model)} - 0.5}{\text{AUC(base model)} - 0.5} - 1 ) \times 100\%.
\end{equation}

\subsection{Results (RQ1)}
We compare our MWUF with three different groups of approaches to testify the effectiveness. We conduct experiments on three datasets and evaluate the mean results over ten runs. The results are shown in Table~\ref{tab:results}, and we list three kinds of MWUF, including MWUF(Wide \& Deep), MWUF(GRU4Rec), MWUF(AFN), which can represent different kinds of models (traditional deep model, sequential model, high-order interaction model). The experimental results further reveal several insightful observations.

\textbf{The effectiveness of SOTA deep CF method.} AFN is a recent deep CF method, which achieves SOTA performance in the regular recommendation scene~\cite{cheng2019adaptive}. We can find that AFN also largely outperforms the popular deep CF methods (Wide \& Deep, PNN, DCN) in the cold-start setting. AFN even outperforms the cold-start methods (DropoutNet, MetaEmb, MeLU) on some tasks. Thus, we think exploring structure design would be a feasible solution for the cold-start problem.

\textbf{The effectiveness of various kinds of cold-start approaches.} The three cold-start baselines can represent three group methods, and our MWUF is another kind of method. We can find that various kinds of cold-start methods are effective to solve the cold-start problem. MeLU has unsatisfying results on Taobao Display AD dataset, and the main reason would be that MeLU is designed to personalize parameters for users, which could be useless in the advertising scene. 

\textbf{The effectiveness of sequential models.} Sequential models focus on effectively exploiting sequential interactions, which is also a challenge in the cold-start problem. We can find that the sequential models have an unsatisfying performance in the cold phase, and the main reason would be that no interaction could have a negative impact on the sequential part of these models. On the contrary, by effectively exploiting the limited interactions, these models achieve remarkable performance on the warm-up phases, which demonstrates the effectiveness of them. While most cold-start researches ignore the sequential methods, we think sequential recommendation is an explorable direction to improve the cold-start performance.


\textbf{Recommendation performance on different phases.} For most methods, the performance in later stages is better, e.g., warm-c is later than warm-b. In the later stage, new (cold) items have more interactions, and the approaches can model them better. Thus, the recommendation performance is better. In other words, with more data, the performance is better. In addition, the improvement (RelaImpr) of cold-start methods decreases in later stages. With more data, the gap between cold ID embedding and the deep model becomes smaller. Thus, in the later stages, the cold-start methods which mainly focus on the gap would have less improvement in recommendation performance.

\textbf{The effectiveness of MWUF.} With different datasets, on most tasks, MWUF outperforms most compared methods, which demonstrates the effectiveness of MWUF. Especially, with the results of the t-test, we can find that MWUF could outperform the best baseline significantly in most scenarios. The improvement mainly comes from the warm ID embeddings of items.








\subsection{Generalization Experiments (RQ2)}
Since both MWUF and MetaEmb focus on the improvement of items' representations, we compare these two methods in more scenarios. While both MWUF and MetaEmb are general framework, we apply MWUF and MetaEmb upon six different base models with three datasets to testify the compatibility, including deep CF methods Wide \& Deep, PNN, DCN, AFN; sequential approaches GRU4Rec, DIN. The results are shown in Figure~\ref{fig:generalization}, the solid lines are averaged scores and the bands are standard deviations over ten runs. We have the following findings from the results.

\textbf{Compatibility.} Both MWUF and MetaEmb can be applied upon various base models. With different base models, the two methods are effective to improve the recommendation performance for new items in both cold-start and warm-up phases. Thus, both the two models have satisfying compatibility.

\textbf{The effectiveness of MWUF.} On most tasks, MWUF can achieve satisfying performance. On the one hand, with various base models, the generalized MWUF can constantly achieve the best results. On the other hand, as the cold-start problem is highly challenging, the achieved AUC is good enough to testify the effectiveness of generalized MWUF in real-world scenarios. From the results, it is clear to see that MWUF adapts the new items faster than MetaEmb, which demonstrates that only exploiting pre-trained embeddings is not sufficient to achieve fast adaptation.

\textbf{Discussion.} Compared with base models, in the cold stage, MWUF and MetaEmb have a more significant improvement on MovieLens and CIKM2019 datasets. We think item ID embeddings could be more important in the two datasets. Especially, in the advertisement dataset, MetaEmb has similar performance to the base models, which indicates that advertisement recommendation (Taobao Display Ad) could be more dependent on other feature embeddings. 

An interesting finding is that the sequential base models (GRU4Rec, DIN) achieve similar performance to MWUF and MetaEmb on CIKM2019 dataset. To explain, recall that CIKM2019 is an E-commerce dataset to predict the purchasing behavior, so the users buying the same item could have a strong similarity. Thus, the sequential feature would dominate the recommendation performance.






\begin{figure}[t!]
\centering
\begin{minipage}[b]{1\linewidth}
\centering
\includegraphics[width=.85\linewidth]{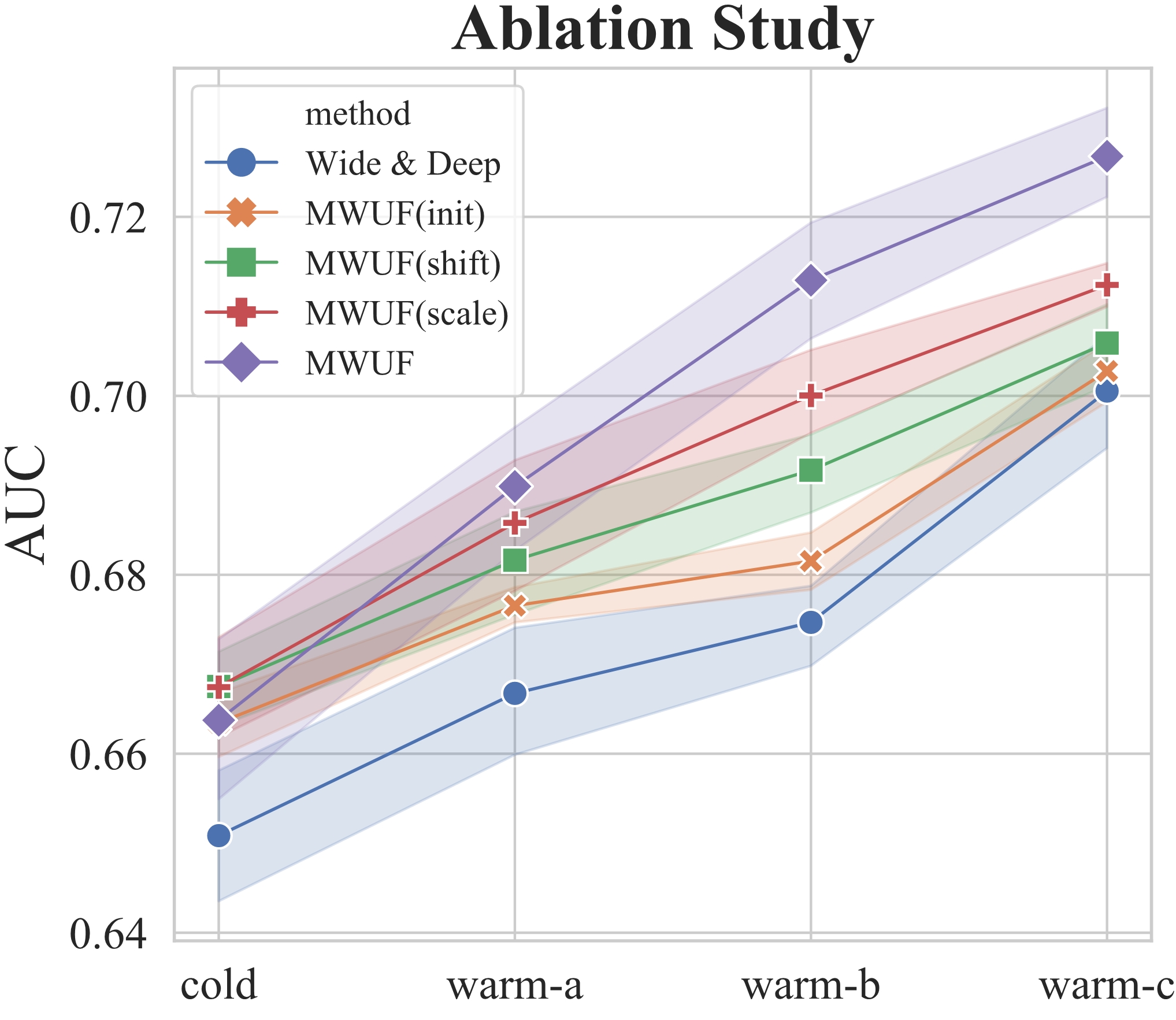}
\end{minipage}
\caption{Ablation Study on CIKM2019 EComm AI dataset. The solid lines are averaged scores and the bands are standard deviations over ten runs.}\label{fig:ablation}
\end{figure} 

\subsection{Ablation Study (RQ3)}
To demonstrate how each component contributes to the overall performance, we now present an ablation test on our MWUF. We conduct ablation study on the CIKM2019 EComm AI data by evaluating several models based on Wide \& Deep: (1) MWUF(init): utilizing average embeddings of all existing items to initialize new item ID embeddings without the two meta networks; (2) MWUF(shift): training with only Meta Shifting Network based on MWUF(init); (3) MWUF(scale): training with only Meta Scaling Network based on MWUF(init); (4) MWUF: the overall framework. The results of ablation study are shown in Figure~\ref{fig:ablation}. MWUF(init) outperforms Wide \& Deep, which demonstrates the common initialization is help for both the cold-start and warm-up phases. MWUF(shift) and MWUF(scale) achieve better performance than MWUF(init), which shows the two meta networks are designed reasonably. The overall MWUF achieves the best results, which demonstrates all parts are in harmony with each other,  forming an effective solution for the cold-start problem in both cold-start and warm-up stages.

An interesting finding is that MWUF(scale) achieves better performance than MWUF(shift), which demonstrates that Meta Scaling Network is more effective than Meta Shifting Network. To explain, recall that the role of the two meta networks: Meta Scaling Network learns to transform the cold ID embedding into warm ID embedding, and Meta Shifting Network learns to enhance the embedding. In other words, the role of the generated scaling function is feature transformation, and the role of the predicted shifting function can be seen as utilizing another representation of the item to enhance the ID embedding. Intuitively, feature transformation could have a larger improvement than representation combination.



\begin{figure}[t!]
\centering
\begin{minipage}[b]{1\linewidth}
\centering
\includegraphics[width=.85\linewidth]{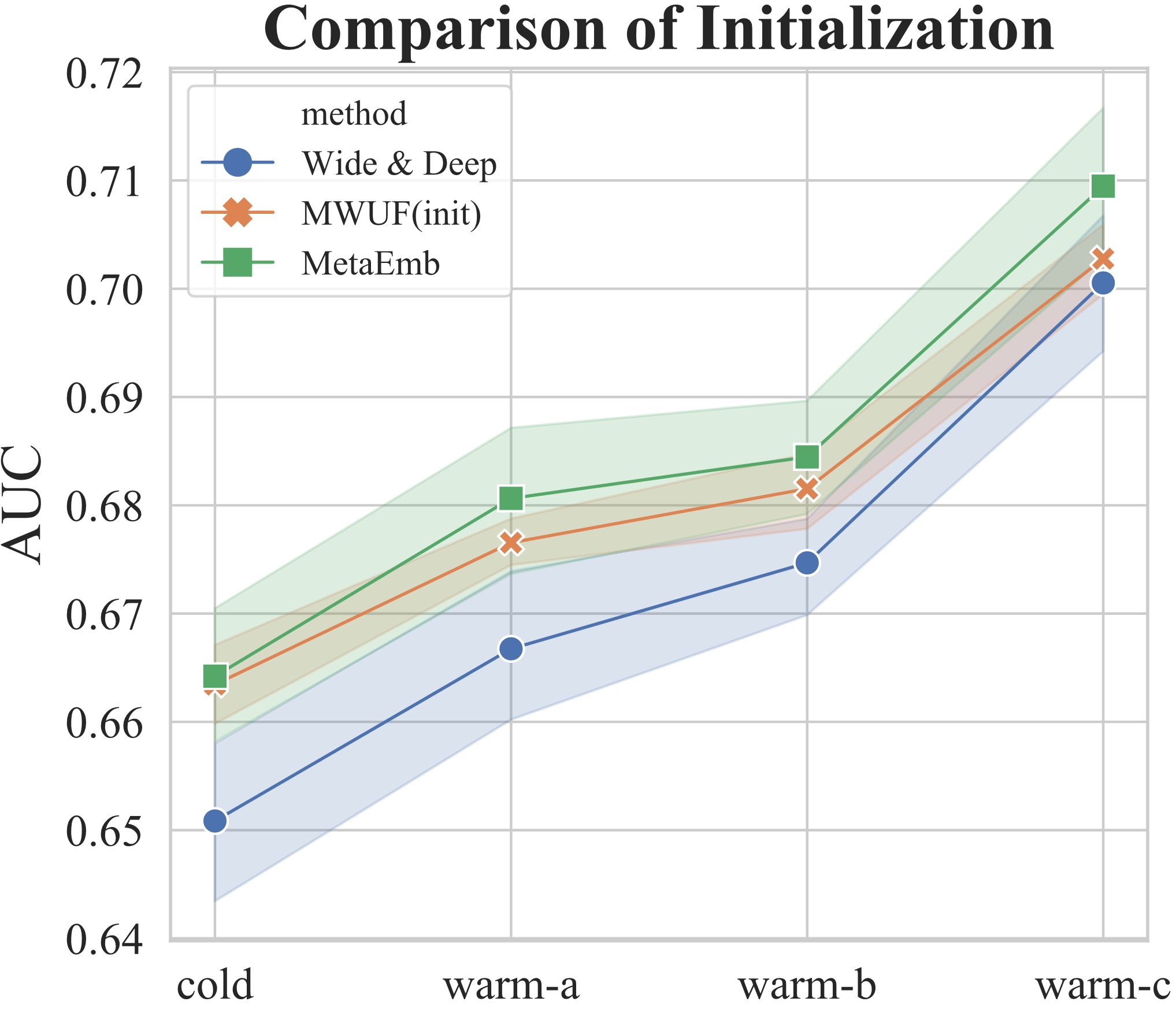}
\end{minipage}
\caption{Comparison of Initialization on CIKM2019 EComm AI dataset. The solid lines are averaged scores and the bands are standard deviations over ten runs.}\label{fig:initialization}
\end{figure} 

\subsection{Comparison of Initialization (RQ4)}
In this subsection, we want to copmpare the influence of different intial methods. We adopt the popular Wide \& Deep as the base model. (1) Randomly initial embedding is usually adopted in industry, and we denote randomly initialization as Wide \& Deep.
(2) MWUF also exploits a simple common initialization method, and we denote the initialization method as MWUF(init) without two meta networks.
(3) MetaEmb utilizes a generator to generate a good ID embedding from item features for each new item, and the generated embedding can be utilized as the initial embedding. We compare the three initialization methods on CIKM2019 EComm AI dataset, and the results are shown in Figure~\ref{fig:initialization}. We can find that both MetaEmb and MUWF(init) outperform Wide \& Deep, which demonstrates a good initial ID embedding is helpful for both cold-start and warm-up phases. In addition, in the cold-start phase, MUWF(init) can obtain comparable results with MetaEmb. The reason would be that the initialization of MetaEmb is learned from side information, and the side information has been exploited repeatedly. Therefore, the function of both MetaEmb and MUWF(init) is to remove the negative influence of random initialization.  In the warm-up phase, MetaEmb outperforms MUWF(init), which demonstrates the specific initialization (side information can represent cluster) is better for common initialization. 


%% file: Conclusion.tex
\section{Conclusion}
In this paper, we studied the cold-start problem. Since the new items only have limited samples, it is hard to train their reasonable ID embeddings to fit the recommendation model. To tackle such problem, we proposed Meta Scaling and Meta Shifting Networks to warm up cold ID embeddings. In detail, the Meta Scaling Network can produce a scaling function for each item to transform the cold ID embedding into warm feature space which fits the model better. Besides, Meta Shifting Network is able to produce a shifting function that can produce stable embeddings from the noisy embeddings. Based on the two meta networks, we proposed the Meta Warm Up Framework (MWUF). Note that the proposed MWUF is a general framework that can be applied upon various existing models using embedding techniques. Finally, we conducted extensive experiments on three real-world datasets to validate the effectiveness and compatibility of our proposed models.